\documentclass[sort&compress,preprint,3p]{elsarticle}
\journal{Annals of Physics}

\usepackage{amsmath}
\usepackage{amssymb}
\usepackage{color}
\usepackage{bm}
\begin{document}

\begin{frontmatter}
\title{Logical inference approach to relativistic quantum mechanics:\\ derivation of the Klein-Gordon equation}
\author[nij]{H.C.~Donker\corref{cor1}}
\ead{h.donker@science.ru.nl}
\author[nij]{M.I.~Katsnelson}
\author[gro]{H.~De Raedt}
\author[jul]{K.~Michielsen}
\cortext[cor1]{Corresponding author}
\address[nij]{Radboud University, Institute for Molecules and Materials, Heyendaalseweg 135, NL-6525AJ Nijmegen, The Netherlands}
\address[gro]{Zernike Institute for Advanced Materials, University of Groningen, Nijenborgh 4, NL-9747 AG Groningen, The Netherlands}
\address[jul]{Institute for Advanced Simulation, J\"ulich Supercomputing Centre, Forschungzentrum J\"ulich, D-52425 J\"ulich, Germany}
\begin{abstract}
{
The logical inference approach to quantum theory, proposed earlier [Ann. Phys. 347 (2014) 45-73], is considered in a relativistic setting.
It is shown that the Klein-Gordon equation for a massive, charged, and spinless particle derives from the combination of the requirements
that the space-time data collected by probing the particle is obtained from the most robust experiment
and that on average, the classical relativistic equation of motion of a particle holds.
}
\end{abstract}


\end{frontmatter}

\section{Introduction}
The inception of quantum theory was one of taking leaps.
This is illustrated by e.g. Schr\"odinger's paper~\cite{SCHR26a} in which he proposes his celebrated wave equation.
In this article~\cite{SCHR26a} Hamilton's principal function $S$ is postulated to take the form
$S =k \ln \psi$ with $k$ a constant and is then substituted in the Hamilton-Jacobi equation (HJE).
Upon variation of the resulting quadratic functional with respect to $\psi$
(which Schr\"odinger later justifies using Huygens' principle~\cite{SCHR26b})
an equation linear in $\psi$, now known as the Schr\"odinger equation, was obtained.
The derivation of the Klein-Gordon equation~\cite{GORD26,FOCK26a,FOCK26b,KUDA26,KLEI27}
is essentially identical to that of the Schr\"odinger equation namely,
an action Ansatz is substituted in the relativistic Hamilton-Jacobi
equation, and after variation of the resulting quadratic functional
with respect to $\psi$, the relativistic analogue of the Schr\"odinger equation is obtained~\cite{GORD26,FOCK26a,FOCK26b,KUDA26,KLEI27}.

Because of the ad hoc assumptions involved in obtaining these equations,
standard quantum mechanics textbooks usually present the formalism of quantum theory as a set of postulates
(see e.g. Refs.~\cite{DIRA57,LAND59, GRIF05, BALL15}) and considerable activity
focuses on eliminating some of these postulates~\cite{BUSC03,CAVE04,RIED07,ZURE03,HARD01,CHIR10,BRUK11,MASA11}.
Instead of starting from a set of postulates, the current work presents an alternative derivation of the
relativistic wave equation based on the principles of logical inference (LI)~\cite{COX46,COX61,TRIB69,JAYN03}.
Specifically, we demonstrate how the Klein-Gordon equation
for a massive, charged and spinless particle
follows from LI based on the analysis of data recorded by a detector, thereby extending
earlier work~\cite{RAED13b, RAED14b, RAED15b, RAED15c} to the relativistic domain.

The key concept in LI is the plausibility~\cite{JAYN03}, a mental construct which quantifies e.g.
the chance that a detection event occurs.
In general, the degree of plausibility is expressed by a real number in the range of 0 and 1~\cite{JAYN03}.
The algebra of LI facilitates plausible reasoning in the presence of uncertainty
in a mathematically well-defined manner~\cite{COX46,COX61,TRIB69,JAYN03}.
In real experiments there is not only uncertainty about the individual detection events
but there obviously is also uncertainty in the conditions under which the experiments are carried out.
Inevitably, the conditions of the experiment will vary whenever the experiment is repeated.
But if the experimental data is to be reproducible,
the experiment must be robust (to be quantified later) with respect to small changes in the conditions under which
the experiment is being performed.
Earlier work has shown that the equations of non-relativistic quantum theory can be obtained by analyzing such robust
experiments~\cite{RAED13b, RAED14b, RAED15b, RAED15c};
most notable are the Schr\"odinger~\cite{RAED14b} and the Pauli equation~\cite{RAED15b}.
Importantly, the requirement that the experiment is to be robust implies that the plausibility must be
viewed as an objective assignment (i.e. conditional probability) rather that a subjective one~\cite{RAED14b}.
The present work extends this approach to the relativistic domain:
it shows how the Klein-Gordon equation~\cite{GORD26,KLEI27} for a massive, charged, and spinless
relativistic particle emerges by an analysis similar to the one employed in Refs.~\cite{RAED14b, RAED15b}.

\section{Logical inference approach}
\subsection{Particle detection experiment}\label{sec:setup}
Consider an experiment in which a particle source and detectors
are located at fixed positions relative to the laboratory reference frame.
The source emits a particle that interacts with one of the detectors
and triggers a detection event that yields data in the form of three spatial coordinates
$\mathbf{r}=(x,y,z)$ of the detector and the clock time $t$ at which the event occurred.
The experiment is considered to be ideal in the sense that every emitted particle triggers one and only
one detector.

The experiment is repeated $N$ times, meaning that we let $N$ particles pass through the detector.
Each time a particle is created, the (laboratory) clock time is reset.
We label the particles and the corresponding data by the index $n = 1 \dots N$
and denote the spatial and temporal resolution by $\Delta_s$ and $\Delta_t$, respectively.
As particle $n$ passes through the detector,
the latter produces a time stamp $t_n$ and a vector of spatial coordinates $\mathbf{r}_n=(x_n,y_n,z_n)$,
which because of the limited resolution, correspond to
the time-bin $j_n = \mathrm{ceiling}(t_n / \Delta_t)$
and space-bin $\mathbf{k}_n = \mathrm{ceiling}(\mathbf{r}_n / \Delta_s)$ where, element-wise,
the function $\mathrm{ceiling}(x)$ returns the smallest integer not smaller than $x$.
In practice the number of time-bins and space-bins is necessarily finite.
Therefore we must have $0\le j_n \le J$ and $(0,0,0)\le \mathbf{k}_n \le \mathbf{K}=(K_x,K_y,K_z)$,
where $J$, $K_x$, $K_y$ and $K_z$ are (large) integer numbers.

The data collected after $N$ repetitions of the experiment is given by the set of quadruples
\begin{equation}
\Upsilon = \left\{ (j_n,\mathbf{k}_n)\; |\; 0\le j_n \le J;\; \mathbf{0}\le \mathbf{k}_n \le \mathbf{K};\; n=1 \dots N \right\} \,,
\end{equation}
or, denoting the total amount of clicks in bin $\mathbf{j}=(j_n,\mathbf{k}_n)$ by $c_\mathbf{j}$,
by the equivalent data set
\begin{equation}
\mathcal{D} = \{ c_\mathbf{j}\;  |\; \sum_\mathbf{j} c_\mathbf{j} = N \} \,.
\end{equation}
Note that at this stage, we have not yet assumed that there is a
relation between the space-time coordinates of the particle and the data set $\mathcal{D}$.

\subsection{Inference probability and Fisher information}
Having specified the measurement scenario, the next step in the LI approach
is to encode the relation between the space-time coordinates of the particle and the $n$-th detection
event $\mathbf{j}_n=(j_n,\mathbf{k}_n)$ through the inference-probability (i-prob)
$P(\mathbf{j}_n |\theta, Z)$ where $\theta$ and $Z$ specify the conditions
under which the experiment is being performed~\cite{RAED14b}.
The i-prob is, at this stage, a necessarily subjective number between zero and one that
expresses the uncertainty with which the $n$-th particle produces the data $\mathbf{j}_n$.
The particle is assumed to be characterized by its own (unknown) clock time $\theta$
measured in a reference frame attached to the particle
The proposition $Z$ represents all other experimental conditions (e.g. applied electromagnetic potentials)
which are considered fixed for the duration of
the experiment but are deemed irrelevant for the problem at hand.

It is common practice to assume, as a first step, that events are independent,
meaning that knowing all earlier and future events, it is impossible to say with certainty what the event will be.
Following this practice, we assume that the $N$ detection events are independent.
Then, according to the algebra of LI~\cite{COX46,COX61,TRIB69,JAYN03},
it follows immediately that the i-prob $P(\Upsilon|\theta,N,Z)$ to observe data set $\Upsilon$ factorizes as
\begin{equation}\label{eq:iid}
P(\Upsilon|\theta,N,Z) = \prod_{n=1}^N P(\mathbf{j}_n |\theta, Z) \,
,
\end{equation}
or, equivalently,
\begin{equation}
P(\mathcal{D} | \theta, N, Z) = N! \prod_{\mathbf{j}} \frac{P(\mathbf{j}|\theta, Z)^{c_\mathbf{j}}}{c_\mathbf{j}!} \, .
\label{pd}
\end{equation}

The salient feature of the experiment considered here is that there is uncertainty about the individual detection events,
that there is uncertainty in the mapping from $\theta$ to the spatial coordinates and the time of the detection events.
However, if the experimental data is to increase our capability to
uncover relations among the observed events at different space-time points,
the experiment must be robust~\cite{RAED14b}.
In the case at hand this means that small changes in the
unknown clock time $\theta$ do not lead to erratic changes in the observed data $\mathcal{D}$,
even though there is no reproducibility on the level of individual events.

It is convenient to express the requirement of robustness as an hypothesis test~\cite{RAED14b}.
The evidence~\cite{TRIB69,JAYN03} $\mathrm{Ev}$ for the hypothesis that $\theta+\epsilon$ produces the data $\mathcal{D}$
relative to the hypothesis that $\theta$ produces the same data
is given by~\cite{TRIB69,JAYN03,RAED14b}
\begin{equation}
\mathrm{Ev} = \ln \frac{P(\mathcal{D} | \theta + \epsilon, N, Z)}{P(\mathcal{D} | \theta, N, Z)} \, . \label{ev0}
\end{equation}
The notion of a robust experiment then translates to the statement
that for all $\theta$ and arbitrary but small $\epsilon$, the evidence $|\mathrm{Ev}|$ should be as small as possible.
In searching for the solution of the global optimization problem,
we exclude the trivial, non-informative experiment for which $P(\mathcal{D} | \theta,N, Z)$ does not depend on $\theta$~\cite{RAED14b}.
Making use of Eq.~(\ref{pd}) and expanding Eq.~(\ref{ev0}) to second order in $\epsilon$ yields
\begin{eqnarray}\label{eq:evidence_expans}
\mathrm{Ev} &=& \sum_{\mathbf{j}} c_\mathbf{j} \ln \frac{P(\mathbf{j}|\theta+\epsilon,Z)}{P(\mathbf{j}|\theta ,Z)}
=
\sum_{\mathbf{j}}  c_\mathbf{j}  \left\{ \epsilon \frac{P^\prime(\mathbf{j}|\theta ,Z)}{P(\mathbf{j}|\theta ,Z)}
 -\right. \left. \frac{\epsilon^2}{2} \left[ \left(\frac{P^\prime(\mathbf{j}|\theta ,Z)}{P(\mathbf{j}|\theta ,Z)}\right)^2
 - \frac{P^{\prime\prime}(\mathbf{j}|\theta ,Z)}{P(\mathbf{j}|\theta ,Z)} \right] \right\} \, ,
\end{eqnarray}
where the primes indicate partial derivatives with respect to $\theta$.

Our goal is now to minimize $|\mathrm{Ev}|$ for all $\theta$ simultaneously.
First note that as $\sum_j P(\mathbf{j}|\theta, Z) = 1$,
all partial derivatives of $\sum_j P(\mathbf{j}|\theta, Z)$ with respect to $\theta$ are zero.
Therefore the first and the third term in Eq.~(\ref{eq:evidence_expans}) vanish if we make the assignment $c_\mathbf{j} = N P(\mathbf{j}|\theta,Z)$.
This is an important result: the criterion of robustness not only enforces
the intuitively obvious assignment $P(\mathbf{j}|\theta,Z)=c_\mathbf{j}/N$
but by doing so, it changes the subjective nature of $P(\mathbf{j}|\theta,Z)$
into an objective, physically measurable quantity (the relative frequency of outcomes).
Thus, it is at this point that the possibility to view the i-prob as a subjective assignment is eliminated~\cite{RAED14b}.

With this assignment, the expression for the evidence becomes%
\begin{equation}
\mathrm{Ev} = -\frac{\epsilon^2 N}{2} \sum_{\mathbf{j}}
\frac{1}{P(\mathbf{j}|\theta, Z)}
\left(\frac{\partial P(\mathbf{j}|\theta, Z)}{\partial \theta}\right)^2
\, ,
\end{equation}
and as $\epsilon$ is arbitrary, we can find the solution of the optimization problem by minimizing
the Fisher information
\begin{equation}
I_F = \sum_{\mathbf{j}}
\frac{1}{P(\mathbf{j}|\theta, Z)}
\left(\frac{\partial P(\mathbf{j}|\theta, Z)}{\partial \theta}\right)^2
\, ,
\label{eq:I_F}
\end{equation}
for all $\theta$ simultaneously.

The basic equations of (relativistic) quantum theory are formulated in terms of  continuous space and time.
Therefore, to derive such equations from a LI approach, it is necessary to take the continuum limit of Eq.~(\ref{eq:I_F}).
This is readily accomplished in the standard manner by letting the temporal resolution $\Delta_t$ and spatial resolution $\Delta_s$ approach zero
while keeping the four dimensions of the four-dimensional volume fixed.
Taking the continuum limit and ignoring irrelevant prefactors, Eq.~(\ref{eq:I_F}) becomes

\begin{eqnarray}
I_F &=& c\int \mathrm{d}^3\mathbf{r}
\int \mathrm{d} t \,
\frac{1}{P(t,\mathbf{r}|\theta, Z)}
\left(\frac{\partial P(t,\mathbf{r}|\theta, Z)}{\partial \theta}\right)^2
\equiv
\int \mathrm{d}^4\mathbf{x}
\frac{1}{P(\mathbf{x}|\theta, Z)}
\left(\frac{\partial P(\mathbf{x}|\theta, Z)}{\partial \theta}\right)^2
,
\label{I_F}
\end{eqnarray}
where $\mathbf{x}=[ct,x,y,z]=[x^0,x^1,x^2,x^3]$ denotes the four-vector of a location in space-time
and $c$ is the speed of light in vacuum.
Strictly speaking, Eq.~(\ref{I_F}) makes a slight abuse of notation:
in the continuum limit $P(\mathbf{x}|\theta, Z)$ is a probability density
whilst $P(\mathbf{x}|\theta, Z)\mathrm{d}\mathbf{x}$ is the corresponding (dimensionless) i-prob.
Henceforth it is assumed that this change of notation is implicitly understood.

\subsection{Special relativity}\label{sec:srt}
The above discussion focussed on the relation between a robust experiment
and the observed data but does not refer to any physical theory yet.
The knowledge or expectation about the physical behavior
enters the LI approach by imposing constraints on the minimization of Eq.~(\ref{I_F}).
Generally speaking, in the absence of uncertainty, we may expect to observe
data that complies with the classical mechanical description,
Thus, in the case at hand, we require that in the absence of uncertainty,
the LI approach yields the results of the special theory of relativity (STR).

Proper time, that is the time measured by a clock at rest, is a central notion in the STR.
In the measurement scenario described above, the i-prob $P(\mathbf{x}|\theta, Z)\mathrm{d}\mathbf{x}$
was already assumed to depend on the proper time $\theta$ of the particle.
In the spirit of the STR, we assume that
\begin{equation}
P(\mathbf{x}|\theta, Z)=P(\tau|\theta, Z)
\, ,
\label{Ptau}
\end{equation}
where $\tau$ ($c^2\tau^2\equiv c^2t^2-x^2-y^2-z^2$) is the proper time in the reference frame of the detector.
In words, we assume that the i-prob to observe a space-time event depends on Lorentz scalars only.
Note that because the clock is being reset with each repetition of the experiment, the proper times
that enter our description are proper time intervals.
In addition, we assume that space-time is homogeneous, meaning that
\begin{equation}
\label{eq:translational_invariance}
P(\tau + \delta | \theta + \delta, Z) = P(\tau | \theta , Z) \, ,
\end{equation}
where $\delta$ is an arbitrary shift of proper time.
A remark about the last assumption may be in order.
In the LI approach the measurement scenario and the observed data are key to formulate the notion of a robust experiment.
Physics on the other hand is about connecting mental pictures, concepts about the nature of the world around us, to the observed data.
Viewed in this light, Eq.~(\ref{eq:translational_invariance}) expresses our expectation that
carrying out the experiment at another point in space-time does not change our mental picture.
From the definition of the derivative, it follows directly from  Eq.~(\ref{eq:translational_invariance}) that
\begin{equation}
\frac{\partial P(\tau|\theta, Z)}{\partial \tau} = -\frac{\partial P(\tau|\theta, Z)}{\partial \theta}
,
\end{equation}
and using this identity, Eq.~(\ref{I_F}) becomes
\begin{equation}
\label{eq:I_F_tau}
I_F =
\int \mathrm{d}^4\mathbf{x}
\frac{1}{P(\mathbf{x}|\theta, Z)}
\left(\frac{\partial P(\mathbf{x}|\theta, Z)}{\partial \tau}\right)^2
=
\int \mathrm{d}^4\mathbf{x}
\frac{1}{P(\tau|\theta, Z)}
\left(\frac{\partial P(\tau|\theta, Z)}{\partial \tau}\right)^2
.
\end{equation}
Recall that the objective of the LI approach is to find $P(\mathbf{x}|\theta, Z)$ that
minimizes $I_F$ for all $\theta$ simultaneously, subject to constraints that we discuss next.

\subsection{Motion of the particle}

In the absence of uncertainty, successive observed detection events map one-to-one
on the relativistic motion of the particle, described by the laws of the STR.
In the LI approach, this limiting case enters through a ``correspondence principle'' in terms of the HJE~\cite{RAED14b,RAED15b}.
This is not a surprise: as mentioned in the introduction, the HJE is one of the key ingredients in the derivation of
the Schr\"odinger~\cite{SCHR26a} and the Klein-Gordon equation~\cite{GORD26, FOCK26a, FOCK26b,KUDA26, KLEI27, MOTZ64}
and it plays a similar role in the LI derivation of these equations.
In the present paper, we do not postulate a HJE but,
in analogy with the derivation of the non-relativistic HJE~\cite{RALS13, RAED15b},
we follow an alternative path and derive the relativistic HJE
for a massive and charged particle from a field description of the four-velocity
$\mathrm{d} \mathbf{x}/\mathrm{d} \tau$.

We start by assuming that there exists a (four-)vector field $\mathbf{U}(\mathbf{x})$ such that
\begin{equation}\label{eq:gauge_def}
\frac{\mathrm{d} x^\mu}{\mathrm{d} \tau} = - U^\mu(\mathbf{x})
\, ,
\end{equation}
Here and in the following, we use the standard co/contra-variant notation and the summation convention
and denote the Minkowski metric by $\eta = \mathrm{diag}(1, -1, -1, -1)$.
Taking the derivative of Eq.~(\ref{eq:gauge_def}) with respect to $\tau$ yields
\begin{equation}\label{eq:derivative_A}
\frac{\mathrm{d}^2
x^\mu}{\mathrm{d} \tau^2} = -\frac{\partial U^\mu}{\partial (ct)}\frac{\mathrm{d}(ct)}{\mathrm{d}\tau}
- \sum_{i=1}^3 \frac{\partial
U^\mu}{\partial x^i} \frac{\mathrm{d} x^i}{\mathrm{d} \tau}
= -\partial^\nu U^\mu \frac{\mathrm{d} x_\nu}{\mathrm{d} \tau}
\, ,
\end{equation}
where $\partial^\mu$ is the shorthand for $\partial/\partial x_\mu$.
As the norm of the four-velocity is $c$, we have $\mathbf{U}^2 \equiv U^\alpha U_\alpha= c^2$
is a constant and hence any derivative thereof is zero.
Therefore we have
\begin{equation}\label{eq:size_A}
\partial^\mu \mathbf{U}^2 = 2 \left( U^0 \frac{\partial U^0}{\partial x_\mu}
- \sum_{i=1}^3 U^i \frac{\partial U^i}{\partial x_\mu} \right)
=2U^\nu \partial^\mu U_\nu
= 0
\,.
\end{equation}
Substitution of Eq.~(\ref{eq:gauge_def}) into Eq.~(\ref{eq:size_A}) yields
\begin{equation}\label{eq:size_A_2}
(\partial^\mu U^\nu) \frac{\mathrm{d} x_\nu}{\mathrm{d}\tau} = 0 \, .
\end{equation}
From Eqs.~(\ref{eq:derivative_A}) and (\ref{eq:size_A_2}) it then follows that
\begin{equation}\label{eq:lorentz} \frac{\mathrm{d}^2
x^\mu}{\mathrm{d} \tau^2} = \left( \partial^\mu U^\nu -\partial^\nu U^\mu \right)
\frac{\mathrm{d} x_\nu}{\mathrm{d} \tau}
= F^{\mu\nu}\frac{\mathrm{d} x_\nu}{\mathrm{d} \tau}
\, .
\end{equation}
Equation~(\ref{eq:lorentz}) has the same form as the Lorentz force equation
of a particle moving in an electrodynamic field~\cite{JACK99}
if we identify $F^{\mu\nu}$ with the field-strength tensor of electrodynamics.
In order that this identification makes sense,
it is necessary to assume that the particles in the particle detection experiment
are massive and charged.

If $S=S(\mathbf{x})$ represents a scalar field,
the transformation $ A^\mu = U^\mu + \partial^\mu S$  yields
\begin{equation}\label{eq:pre_rhje}
(\bm{\partial} S - \mathbf{A})^2 = c^2
\, .
\end{equation}
where we introduced the shorthand notation $(\bm{\partial} S)^2=(\partial_\alpha S) (\partial^\alpha S)$.
As it is the four-velocity d$\mathbf{x}$/d$\tau$ which corresponds to a physically relevant quantity,
imposing gauge invariance enforces introducing a non-vanishing canonical momentum $p^\mu = \partial^\mu S$
in order keep the norm of the four-velocity fixed to $c$.
Equation Eq.~(\ref{eq:pre_rhje}) is the relativistic HJE in disguise.
Indeed, making use of
$\bm{\partial} S =[
{\partial S}/{\partial x_0},
{\partial S}/{\partial x_1},
{\partial S}/{\partial x_2},
{\partial S}/{\partial x_3}]
=
[{\partial S}/{\partial x^0},
-{\partial S}/{\partial x^1},
-{\partial S}/{\partial x^2},
-{\partial S}/{\partial x^3}]
$
and
$\mathbf{A}=[A^0,A^1,A^2,A^3]=[A_0,-A_1,-A_2,-A_3]$,
introducing the symbols $m$ and $q$ for the mass and charge
of the particle, respectively,
and changing in Eq.~(\ref{eq:pre_rhje}) symbols according to
$\bm{\partial} S \rightarrow(
{\partial S}/{\partial ct},
{\partial S}/{\partial x},
{\partial S}/{\partial y},
{\partial S}/{\partial z})/m
$
and
$\mathbf{A}\rightarrow q(\Phi,A_x,A_y,A_z)/m$
(where $\Phi$ and $(A_x,A_y,A_z)$ are the usual scalar and vector potential, respectively~\cite{JACK99}),
we find
\begin{equation}\label{eq:rhje}
\left(\frac{\partial S(\mathbf{x})}{\partial ct}-\frac{q}{c}\Phi(\mathbf{x})\right)^2
=\left(\frac{\partial S(\mathbf{x})}{\partial x}+\frac{q}{c}A_x(\mathbf{x})\right)^2
+\left(\frac{\partial S(\mathbf{x})}{\partial y}+\frac{q}{c}A_y(\mathbf{x})\right)^2
+\left(\frac{\partial S(\mathbf{x}|\theta, Z)}{\partial z}+\frac{q}{c}A_z(\mathbf{x})\right)^2
+m^2c^2
\;,
\end{equation}
which is the relativistic HJE for a charged, massive particle in an electromagnetic field~\cite{KLEI27, GORD26}.

\subsection{Derivation of the Klein-Gordon equation}
As a first step, it is expedient to write Eq.~(\ref{eq:I_F_tau}) in an alternative form by
noting that $c \tau = \sqrt{\eta_{\mu \nu} x^\mu x^\nu}$ implies $\partial^\alpha \tau = x^\alpha/(c^2\tau)$
such that
\begin{equation}
\eta_{\mu\nu} \left( \partial^\mu P(\tau|\theta, Z)\right)
\left( \partial^\nu P(\tau|\theta, Z)\right) =
\left(\frac{\partial P(\tau|\theta, Z)}{\partial \tau}\right)^2
\eta_{\mu\nu} (\partial^\mu \tau) (\partial^\nu \tau)
= \frac{1}{c^2}
\left(\frac{\partial P(\tau|\theta, Z)}{\partial \tau}\right)^2 \,
,
\end{equation}
and hence Eq.~(\ref{eq:I_F_tau}) can be written as
\begin{equation}\label{eq:F}
I_F = c^2\int \mathrm{d}^4\mathbf{x}
\frac{1}{P(\tau|\theta, Z)}
\left(\bm{\partial} P(\tau|\theta, Z)\right)^2
.
\end{equation}

The general guiding principle of the LI approach is that the experiment that yields the
most robust data is described by the probability density $P(\mathbf{x}|\theta, Z)$ that minimizes $I_F$
for all $\theta$ simultaneously, subject to additional constraints that are deemed
relevant to the experiment at hand~\cite{RAED14b}.
In the present case, we require that the description is compatible with the special theory of relativity.
For a massive, charged particle and in the absence of uncertainty,
the latter requirement implies that the classical, relativistic HJE Eq.~(\ref{eq:pre_rhje}) should hold.
We can inject this requirement into the LI approach by considering the functional
\begin{equation}\label{eq:minimize}
F = c^2 \int \mathrm{d}^4\mathbf{x}
\left\{
\frac{\left(\bm{\partial} P(\tau|\theta, Z)\right)^2
}{P(\tau|\theta, Z)}
+ \lambda \left[ (\bm{\partial} S -\mathbf{A})^2 - c^2\right]P(\tau|\mathbf{\theta}, Z) \right\} \,
,
\end{equation}
where $\lambda$ is a weighting factor that
reflects the importance of the uncertainty and robustness relative to the contribution of the classical dynamics.
It is straightforward to show that the expression Eq.~(\ref{eq:minimize}) is invariant under Lorentz transformations.

Extremization of Eq.~(\ref{eq:minimize}) can be carried out by the standard variational calculus
and yields a set of nonlinear partial differential equations for $P(\mathbf{x}|\theta, Z)$ and $S(\mathbf{x})$.
It is not difficult to show that at an extremum, (i) the value of $F$ does not depend on the value of
the unknown proper time $\theta$ and that (ii) the value of $F$ is zero, independent of $\lambda$.
The latter result implies that the extrema describe situations in which the
uncertainty about the detection events is perfectly balanced by the certainty that
the classical HJE describes the motion of the observed detection events.

It is now expedient to write Eq.~(\ref{eq:minimize}) more explicitly as
\begin{eqnarray}
F &=& c^2\int \mathrm{d}^4\mathbf{x}
\bigg\{
\frac{1}{P(\mathbf{x}|\theta, Z)}\bigg[
\left(\frac{\partial P(\mathbf{x}|\theta, Z)}{\partial ct}\right)^2
-\left(\frac{\partial P(\mathbf{x}|\theta, Z)}{\partial x}\right)^2
-\left(\frac{\partial P(\mathbf{x}|\theta, Z)}{\partial y}\right)^2
-\left(\frac{\partial P(\mathbf{x}|\theta, Z)}{\partial z}\right)^2
\bigg]
\nonumber \\
&&
\hbox to1.5cm{}
+ \lambda \bigg[
\left(\frac{\partial S(\mathbf{x})}{\partial ct}-A^0(\mathbf{x})\right)^2
-\left(\frac{\partial S(\mathbf{x})}{\partial x}+A^1(\mathbf{x})\right)^2
-\left(\frac{\partial S(\mathbf{x})}{\partial y}+A^2(\mathbf{x})\right)^2
\nonumber \\
&&
\hbox to2.5cm{}
-\left(\frac{\partial S(\mathbf{x})}{\partial z}+A^3(\mathbf{x})\right)^2
- c^2
\bigg]P(\mathbf{x}|\theta, Z) \bigg\} \,
.
\label{F}
\end{eqnarray}
We do not know of any direct method to solve the nonlinear set of equations
that results from searching for the extrema of Eq.~(\ref{F}) but, by analogy with
the non-relativistic case, we may consider a quadratic functional of a complex-valued field
$\varphi(\mathbf{x})$ and use the polar representation of this field
to construct the corresponding functional in terms of this representation~\cite{RAED14b,RAED15b,MADE27}.

To this end, consider the quadratic functional
\begin{eqnarray}
Q &=& 4c^2\int \mathrm{d}^4\mathbf{x}
\left\{
\left[
\frac{\partial \varphi^\ast(\mathbf{x})}{\partial ct}+\frac{i\sqrt{\lambda}}{2c}A^0(\mathbf{x})\varphi^\ast(\mathbf{x})
\right]
\left[
\frac{\partial \varphi(\mathbf{x})}{\partial ct}-\frac{i\sqrt{\lambda}}{2c}A^0(\mathbf{x})\varphi(\mathbf{x})
\right]
\right.
\nonumber \\
&&
\hbox to1.5cm{}
-\left[
\frac{\partial \varphi^\ast(\mathbf{x})}{\partial x}-\frac{i\sqrt{\lambda}}{2c}A^1(\mathbf{x})\varphi^\ast(\mathbf{x})
\right]
\left[
\frac{\partial \varphi(\mathbf{x})}{\partial x}+\frac{i\sqrt{\lambda}}{2c}A^1(\mathbf{x})\varphi(\mathbf{x})
\right]
\nonumber \\
&&
\hbox to1.5cm{}
-\left[
\frac{\partial \varphi^\ast(\mathbf{x})}{\partial y}-\frac{i\sqrt{\lambda}}{2c}A^2(\mathbf{x})\varphi^\ast(\mathbf{x})
\right]
\left[
\frac{\partial \varphi(\mathbf{x})}{\partial y}+\frac{i\sqrt{\lambda}}{2c}A^2(\mathbf{x})\varphi(\mathbf{x})
\right]
\nonumber \\
&&
\hbox to1.5cm{}
\left.
-\left[
\frac{\partial \varphi^\ast(\mathbf{x})}{\partial z}-\frac{i\sqrt{\lambda}}{2c}A^3(\mathbf{x})\varphi^\ast(\mathbf{x})
\right]
\left[
\frac{\partial \varphi(\mathbf{x})}{\partial z}+\frac{i\sqrt{\lambda}}{2c}A^3(\mathbf{x})\varphi(\mathbf{x})
\right]
- \frac{\lambda c^2}{4}  \varphi^\ast(\mathbf{x})\varphi(\mathbf{x})
\right\} \,.
\nonumber \\
\label{Q}
\end{eqnarray}
Substituting the polar representation
\begin{equation}
\varphi(\mathbf{x}) = \sqrt{P(\mathbf{x}|\theta, Z)} e^{i\sqrt{\lambda} S(\mathbf{x})/2}
\, ,
\label{PHI}
\end{equation}
in Eq.~(\ref{Q}) yields $Q=F$.
Equations for the extrema of the functional $Q$ can be found
by variation with respect to $\varphi^\ast(\mathbf{x})$, yielding
the linear partial differential equation
\begin{equation}\label{KG0}
\left[
\frac{\partial }{\partial ct}+i\frac{\sqrt{\lambda}}{2c}A^0(\mathbf{x})\right]^2 \varphi(\mathbf{x})
=
\left\{
\left[
\frac{\partial }{\partial x}-i\frac{\sqrt{\lambda}}{2c}A^1(\mathbf{x})\right]^2
+
\left[
\frac{\partial }{\partial y}-i\frac{\sqrt{\lambda}}{2c}A^2(\mathbf{x})\right]^2
+
\left[
\frac{\partial }{\partial z}-i\frac{\sqrt{\lambda}}{2c}A^3(\mathbf{x})\right]^2
- \frac{\lambda c^2}{4}
\right\}
\varphi(\mathbf{x})
\;,
\end{equation}
which has the same mathematical structure as the KG equation~\cite{FESH58}.
This can be made more explicit by changing symbols according to
$\mathbf{A}\rightarrow q(\Phi,A_x,A_y,A_z)/m$
and $\lambda=4m^2/\hbar^2$, yielding
\begin{equation}\label{KG0z}
\frac{1}{c^2}
\left[
i\hbar\frac{\partial }{\partial t}-q\Phi(\mathbf{x})\right]^2 \varphi(\mathbf{x})
=
\left\{
\left[
\frac{\hbar}{i}\frac{\partial }{\partial x}-\frac{q}{c}A_x(\mathbf{x})\right]^2
+
\left[
\frac{\hbar}{i}\frac{\partial }{\partial y}-\frac{q}{c}A_y(\mathbf{x})\right]^2
+
\left[
\frac{\hbar}{i}\frac{\partial }{\partial z}-\frac{q}{c}A_z(\mathbf{x})\right]^2
+ m^2c^2
\right\}
\varphi(\mathbf{x})
\;.
\end{equation}
Obviously, the weighting factor $\lambda=4m^2/\hbar^2$ cannot be determined
on the basis of logic only but has to follow from a comparison of the outcome
of calculations based on Eq.~(\ref{KG0}) with experimental data.
It is of interest to enquire to what extent Eq.~(\ref{KG0z})
allows us to infer from the observed data properties of the massive charged particles.
The speed of light in vacuum $c$ certainly does not depend on
the properties of the massive charged particle.
Then, from Eq.~(\ref{KG0z}), it is immediately clear that its solutions
are invariant under the transformation
$\hbar \rightarrow \hbar \xi$,
$q \rightarrow q \xi$, and
$m \rightarrow m \xi$.
Hence, from the observed data we may be able to determine two but not three of the constants
that appear in Eq.~(\ref{KG0z}).
For instance, by a suitable redefinition of the units of mass and charge,
$\hbar$ can be eliminated from Eq.~(\ref{KG0z})~\cite{RALS13b}.

In practice, instead of solving the set of nonlinear equations in terms of
$P(\mathbf{x}|\mathbf{\theta}, Z)$ and $S(\mathbf{x})$ that result
from minimizing Eq.~(\ref{F}),
it is much easier to first solve Eq.~(\ref{KG0}) and then use Eq.~(\ref{PHI}) to find
$P(\mathbf{x}|\theta, Z)=\varphi(\mathbf{x})^\ast\varphi(\mathbf{x})$.
It is important to recognize that the LI approach gives us the
probability for observing a space-time event $\mathbf{x}$ but does not yield
an estimate of the proper time of the particle $\theta$.
The latter was and remains unknown.
The LI approach suggests that the wave function $\varphi(\mathbf{x})$ is only a mathematical
vehicle, be it an extraordinarily useful one, to transform a set of nonlinear partial differential equations
into a linear set of partial differential equations.
The interplay of the two real quantities $S(\mathbf{x})$ and $P(\mathbf{x}|\theta, Z)$
which account for respectively, the classical relativistic physics and the uncertainty on the collected data,
can be disentangled through the use of single complex wave function.
But, as a mathematical tool, the wave function does not need an interpretation:
it is $P(\mathbf{x}|\theta, Z)$ that is directly linked to the observed events.

\section{Discussion}

We have shown how the Klein-Gordon equation for massive, charged particles derives from logical inference applied
to experiments for which the observed events are independent
and for which the frequency distribution of these events is robust with respect to small changes of
the conditions under which experiments are carried out.
The present derivation is a logical generalization of earlier work~\cite{RAED13b,RAED14b,RAED15b,RAED15c}
to the relativistic domain, the fundamental difference being that the measured time is subject to uncertainty.

Obviously, the transition from non-relativistic to relativistic quantum theory is expected to bring in some radically new features.
Landau and Peierls~\cite{LP31} pointed out that in relativistic quantum theory the particle position cannot be measured
with an accuracy higher than its Compton wavelength.
Measuring the position of an electron with an accuracy higher than its Compton wavelength
requires an energy that exceeds the threshold for the creation of electron-positron pairs~\cite{FESH58},
rendering meaningless the question which of the electrons is the original one.
Therefore, there is a common believe that relativistic quantum theory cannot be a theory of individual particles
but it must be a {\it field theory} for a non-constant number of particles~\cite{PESK95,ZEE10}.
The requirement of a field theory description is also linked to the fact that the charge density of the Klein-Gordon
equation is not positive definite, as mentioned by Dirac~\cite{DIRA28} and also stressed by
Feshbach and Villars~\cite{FESH58}.
This is due to the second order time derivative in the Klein-Gordon equation and indicates
that the wave function describes in fact two degrees of freedom instead of one~\cite{FESH58}.

It is worth noting that there is no mention of the direction of time in the logical inference approach. Indeed, when we derived Eq.~(\ref{eq:I_F_tau}) we allowed both $\theta < \tau$ and $\theta > \tau$. This looks unusual from the point of view of single-particle quantum mechanics (see, however, a discussion of time (a)symmetry in Refs.~\cite{CRAM86,AHAR10}). Within relativistic quantum mechanics it seems more natural. As was suggested by Wheeler one might interpret anti-particles as particles with the sign of the proper time reversed, i.e. as if the particles are moving backward in time~\cite{FEYN98}, and, at least, for non-interacting particles this interpretation seems to be possible. In the measurement scenario analyzed here, there is no way to discern the absorption of particles from the emission of anti-particles. Both types of events contribute equally well to the detection counts. This relates to the measurement scenario where we make no distinction between detection events for which $\tau > \theta$ and detection events for which $\tau < \theta$; causality is not a prerequisite in the derivation presented here.

Naturally one might ask how to extent this approach to particles with non-zero spin (e.g., Dirac equation). We leave this challenging program for future research.

\section*{Acknowledgments}
MIK and HCD acknowledges financial support by the European Research Council, project 338957 \\FEMTO/NANO.

\section*{References}
\bibliographystyle{elsarticle-num}
\bibliography{KGeq}
\end{document}